# Physiological Biomarkers of State Anxiety: Feature-Based Algorithm Development and Validation Using Electrooculography and Electrodermal Activity During a Cold Pressor Test

Jadelynn Dao[1], Ruixiao Liu[1], Sarah Solomon[2], Samuel Solomon[1*]

Affiliations.

[1]Andrew and Peggy Cherng Department of Medical Engineering, Division of Engineering and Applied Science, California Institute of Technology, Pasadena, CA, USA.

[2]Dartmouth Hitchcock Medical Center and Clinics, Adult Psychiatry Residency, Lebanon, NH, USA.

*Corresponding author. Email: ssolomon@caltech.edu.

## Abstract

### Background

Anxiety has become a significant health concern affecting mental and physical well-being, with state anxiety—a transient emotional response—linked to adverse cardiovascular and long-term health outcomes. Traditional physiological monitoring lacks the contextual sensitivity needed to assess anxiety in real time. Electrooculography (EOG) and electrodermal activity (EDA), two bio signals measurable by wearables, offer promising avenues for identifying biomarkers of state anxiety (s-anxiety) in naturalistic environments.

### Objective

This study aims to identify novel biomarkers of state anxiety using electrooculography (EOG) and electrodermal activity (EDA) signals collected in real-world conditions. We further explore how non-invasive wearable technology can enable real-time monitoring of physiological responses during induced stress, focusing on distinguishing true anxiety-related signals from artifacts in noisy environments.

### Methods

Our study presents two datasets: 1) EOG signal blink identification dataset BLINKEO, containing both true blink events and motion artifacts, and 2) EOG and EDA signals dataset EMOCOLD, capturing physiological responses from a Cold Pressor Test (CPT). From analyzing blink rate variability, skin conductance peaks, and associated arousal metrics, we identified multiple new anxiety-specific biomarkers. SHapley Additive exPlanations (SHAP) were used to interpret and refine our model, enabling a robust understanding of the biomarkers that correlate strongly with state anxiety.

## Results

BLINKEO feature analysis achieved a classification accuracy of 98.17% and F1 score of 0.8734 in distinguishing blinks from noise. In EMOCOLD, survey results confirmed elevated anxiety and affectivity during CPT, which normalized during recovery. SHAP analysis revealed that specific EDA features (e.g., Hjorth Activity, Spectral Entropy) and EOG features (e.g., Opening Phase Energy, Signal Height) consistently contributed to accurate predictions of state anxiety and affectivity. Contextual combinations of features outperformed single-feature analyses, revealing relationships critical for robust biomarker identification.

## Conclusions

These results suggest that a combined analysis of EOG and EDA data offers significant improvements in detecting real-time anxiety markers, underscoring the potential of wearables in personalized health monitoring and mental health intervention strategies. This work contributes to the development of context-sensitive models for anxiety assessment, promoting more effective applications of wearable technology in healthcare.

# Introduction

Despite being a short-term response, state anxiety (s-anxiety) has emerged as a significant factor impacting long-term health outcomes. Researchers have linked sustained s-anxiety with adverse cardiovascular effects [1], underscoring its profound effects on mental and physical health. Approximately 23.1% of American adults experience some form of diagnosable mental disorder [2] and 74% of American adults reported experiencing stress-related health issues within a given month [3], illustrating the widespread impact of anxiety-induced stress. Reliable biomarkers are essential for capturing the complexities of s-anxiety, enabling more precise and effective models.

Non-invasive wearable technology has the potential to transform health monitoring by continuously capturing physiological data through real-time sensor measurements [4]. These devices collect a broad array of metrics, yielding critical insights into the body's responses to anxiety. The ability to seamlessly collect large amounts of health-related data opens new ways to study and build an understanding of the onset and progression of anxiety, enabling more effective interventions and advancing our knowledge of human health. Identifying reliable biomarkers of s-anxiety offers a promising pathway to real-time health monitoring using wearable biosensors that can detect subtle physiological changes not immediately obvious in raw signal data.

The cold pressor test (CPT) is a widely utilized experimental method for studying anxiety responses in controlled settings. Participants immerse their hand in ice-cold water (0–4°C), eliciting a sympathetic nervous system response. This test reliably induces physiological markers of anxiety [5], [6], [7], such as increased heart rate and sweat production. Other techniques, such as public speaking simulations and mental arithmetic tasks [8], also provoke anxiety and can be used to identify reliable biomarkers.

Physiological responses to s-anxiety and arousal have been extensively documented, revealing clear links between emotional states and indicators such as blink rate variability [9] and stress-induced sweating [10] The two-factor model of emotion, developed by Schachter and Singer [11], suggests that emotions arise from physiological arousal and subsequent cognitive interpretation. This model underscores that physiological responses are interpreted within a contextual framework, which are further hidden in indirect biomarkers for specific emotional experiences. For instance, fatigue, which affects the blink conditions, can intensify physiological arousal, directly impacting how the brain interprets anxious states. Such contextual cues are crucial for understanding s-anxiety in real-world settings, but they are often filtered out or controlled for in existing studies. Electrodermal activity (EDA) is a common measure of physiological arousal, but its reliability in depression research remains debated. Some studies report reduced EDA responses in individuals with major depressive disorder, suggesting impaired autonomic reactivity [12] and emotional hypo-responsiveness [13]. However, conflicting findings point to variability due to factors like medication use and methodological differences [14], emphasizing the need for further research on the relationship between physiological signals and emotional states.

Wearable devices offer a way to contextualize these arousal states dynamically. Through advanced human-machine interfaces, wearables can monitor how individuals respond to their environments, integrating data on physical responses to build a richer understanding of s-anxiety. There is growing interest in using non-invasive wearables to collect richer biomarker data for mental health study [15], [16], interpreting physiological responses in respect to real-time contextual cues and providing a more comprehensive view of emotional states.

Research shows that blink rates tend to increase under difficult mental tasks or anxiety-provoking situations [17], [18], reflecting activation of the autonomic nervous system. Electrooculography (EOG) captures electrical signals produced by eye movements, allowing for the detection of blink-related patterns. But EOG signals are often filtered out in stress studies to improve clarity of other signals [19], potentially overlooking valuable information related to emotional arousal. Studies suggest that specific components of EOG signals can be analyzed to extract physiological markers of s-anxiety, highlighting the need for further research into EOG biomarkers. Moreover, fatigue—closely associated with emotional arousal—provides an additional avenue for understanding anxiety through EOG features [20], [21]. Studies examining EOG signals in the context of drowsiness reveal correlations between blink frequency, blink duration, and stages of fatigue [18], highlighting a non-invasive method for tracking emotional arousal over time. Given the interplay between fatigue and anxiety, this relationship prompted our investigation into how fatigue-related features within EOG signals may serve as indirect indicators of anxiety, offering new opportunities for nuanced and comprehensive stress monitoring.

Similarly, stress has a pronounced effect on sweat production. Emotional sweating, triggered by the sympathetic nervous system, occurs in response to psychological stressors rather than temperature changes [14] [15]. Electrodermal activity (EDA) is a method that measures changes in skin conductance. Under emotional arousal and stress, body sweats and skin conductance increases. Previous studies often rely on basic features like median values [24] or the phasic component of the EDA signal, focusing on nonspecific skin conductance responses (SCR) to correlate with self-reported s-anxiety [25] scores. In such studies, peaks in the phasic signal exceeding 0.01 μS were counted as responses, and the frequency of these nonspecific SCRs per minute served as the primary measure for physiological s-anxiety. EDA primarily reflects the *magnitude* of emotional arousal without distinguishing between positive and negative affective states [26]. In other words, a high SCR could result from excitement or stress, making it challenging to interpret EDA data as a standalone indicator of anxiety. This underscores the importance of using EDA in combination with other physiological markers [27], such as heart rate variability or blink rate, to gain a more comprehensive picture of an individual's emotional and physiological state. A more methodical exploration of signal characteristics found in EDA and EOG signals reveal nuanced physiological markers that strongly correlate with s-anxiety.

Currently, no widely accepted biomarkers reliably assess anxiety across diverse contexts, highlighting the need for continued exploration. Researchers have tested markers like heart rate variability, skin conductance, and blink rate, but results often vary due to individual differences and contextual influences. While many studies report that depressed patients exhibit reduced EDA responses, indicating diminished autonomic nervous system activity, some research presents conflicting findings. These discrepancies are attributed to variations in study designs, methodologies, and the influence of factors such as antidepressant treatment on EDA measurements [12].

While machine learning models have shown promise in detecting anxiety, their black-box nature limits interpretability, making it difficult to validate findings across diverse populations [28]. By introducing additional context-sensitive biomarkers, we aim to enhance the reliability and transparency of anxiety assessments, making models more applicable to real-world scenarios.

## Objective

In our research, we leverage EOG and EDA data to develop a comprehensive, real-time model of s-anxiety. We have compiled two distinct datasets for this purpose. The first dataset, *BLINKEO*, consists of EOG signal features from samples characterized by peak-like patterns, annotated to differentiate natural blink events from extraneous noise and wire movement artifacts. The second dataset, *EMOCOLD*, contains time-series EOG and EDA signals along with demographic data and stress responses elicited by the Cold Pressor Test (CPT). Using interpretability techniques such as SHAP (SHapley Additive exPlanations), we identify and quantify specific biomarkers within the EOG and EDA data, with a focus on blink rate variability and sweat-related stress indicators. Our approach goes beyond simple anomaly detection by uncovering nuanced, anxiety-specific physiological markers informed by the two-factor model of emotion. This research contributes to a more detailed understanding of stress mechanisms, with the potential to improve mental health interventions and enable personalized, context-specific stress management strategies with wearable technology.

## Description of Question

This research aims to identify reliable, interpretable biomarkers of s-anxiety through EOG and EDA data for real-time stress monitoring.

# Methods

## Blink Identification EOG (BLINKEO) Data Collection

To create the BLINKEO Blink Identification EOG Dataset, EOG data was collected and analyzed to differentiate natural blinks from noise or wire movements. Our setup integrated the AD8232 (Analog Devices), a biopotential amplifier designed to capture physiological signals, which we optimized for measuring EOG activity. To detect vertical eye movements using EOG, one electrode was positioned above the eye and another below it, aligning on the vertical axis. This configuration captures the corneo-retinal potential changes associated with upward and downward eye movements. All trials were conducted on the same two individuals for consistency in signal characteristics. 65 trials involving repeated blinking under controlled conditions where no extraneous movement occurred. Additionally, 19 trials lasting between 30 seconds and 2 minutes were conducted under conditions with no blinking, but with deliberate wire movements introduced by manually adjusting or lightly tugging the electrode leads. These trials provided a baseline for accurately distinguishing noise artifacts from genuine blink events. Table 1 summarizes the characteristics of these trials, including session count, total recording time, and peak detection results before and after filtering.

Table 1

| **Trial Label** | **No. Sessions** | **Total Time (s)** | **No. Peaks Detected** | **No. Peaks After Filtering** |
|---|---|---|---|---|
| Blink | 65 | 12103.14 | 6792 | 4734 |
| Wire Movement | 19 | 2007.75 | 5704 | 203 |

*Table 1* Characteristics of blink and wire movement trials in the Blink Identification Dataset. This table summarizes the number of independent sessions, cumulative recording time, and peak detection results before and after literature-supported blink peaks filtering for both blink and wire movement events.

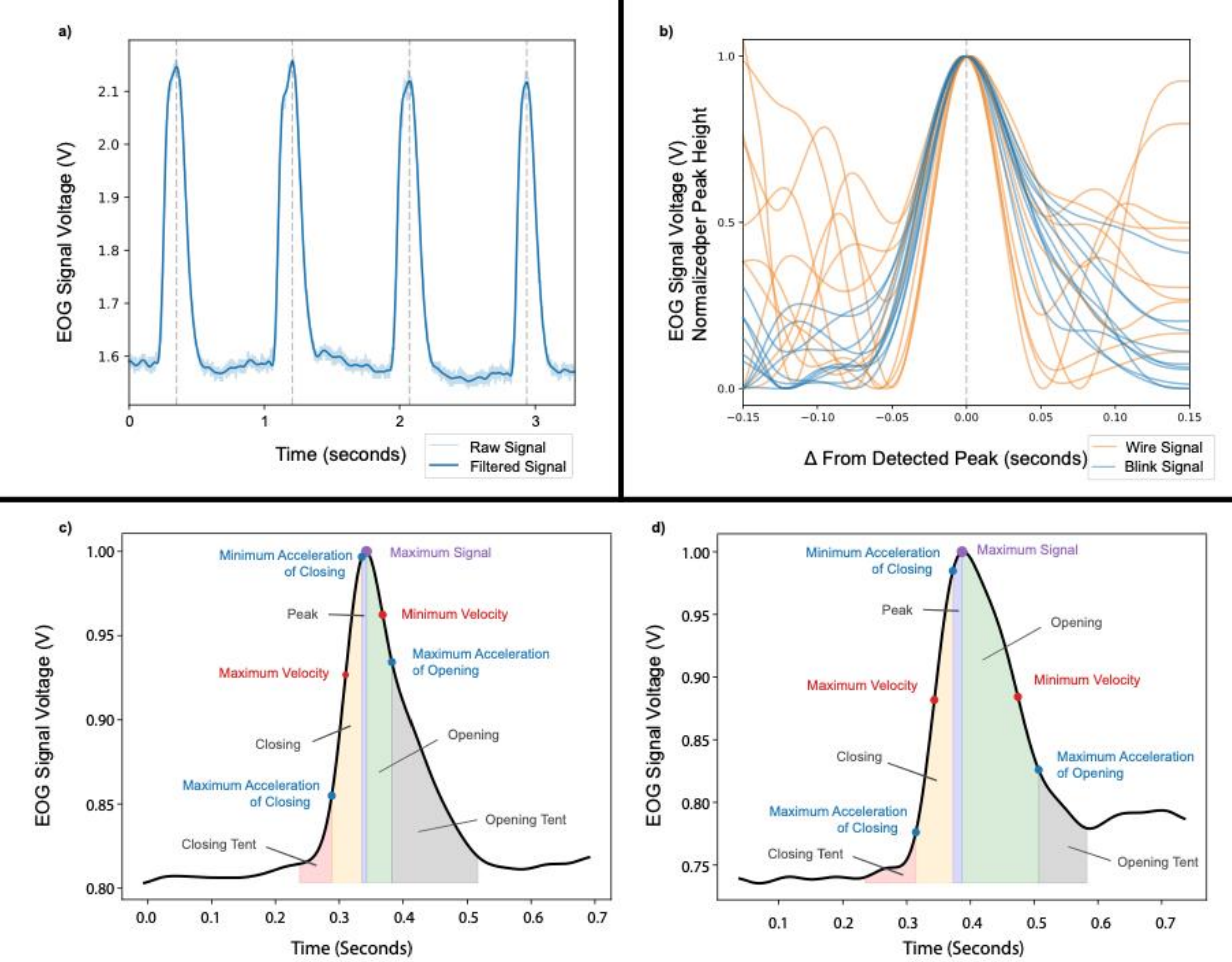

*Figure 1 1a. Blink Peak Examples from the BLINKEO dataset. The grey dotted lines indicate the center of the peak, extracted by the peak detection method outlined in this section. 1b. Blink examples (blue) plotted against wire examples (green), as filtered EOG Voltage signals, normalized per peak between 0 and 1. Peaks are time-aligned by time, in seconds, from the center of peak. Wire signals typically have higher variability. 1c-d. c. A singular blink peak. The purple dot marks the peak of the blink event, while the outer edges of the red and grey shaded sections represent the boundaries used for feature extraction. These boundaries are determined by identifying the nearest minimum on each side of the peak, providing a precise range for analyzing blink characteristics. d. Another example of a blink peak, demonstrating the variability in blink peak shapes observed across recordings. The feature extraction process remains consistent, with boundaries determined by identifying the nearest minima on either side of the peak.*

To preprocess the EOG data, motion artifacts were identified and removed, to make the data suitable for downstream features. A fifth-order low-pass Butterworth filter using the Scipy Signal *butter* function was applied to isolate low-frequency components indicative of meaningful physiological signals. This was followed by a Savitzky-Golay filter using the Scipy Signal *savgol_filter* function for additional smoothing, which preserved essential features while reducing minor signal fluctuations [29].

Peak detection was performed using the Scipy Signal *find_peaks* function, identifying peaks with a prominence exceeding 0.1 with a peak width greater than 0.04 seconds [30] (blinks typically last between 0.1 and 0.4 seconds [31], averaging around 0.25 seconds). To focus on blink-like events, we additionally applied criteria based on established blink characteristics: a maximum peak width of 0.5 seconds and a minimum peak height of 0.05 volts [30]. We compared the signal quality after this initial peak detection with that obtained using conventional blink filtering methods. Traditional filtering techniques frequently overlook subtle blink patterns or introduce artifacts during data cleaning, potentially compromising accuracy. In contrast, a learned-feature approach refines this process by reducing noise and enhancing the precision of true blink identification within the dataset. Figure 1a illustrates examples of detected blink peaks from the BLINKEO dataset, with red dotted lines marking the center of each peak. This figure demonstrates the effectiveness of the peak detection method described in this section, highlighting its ability to accurately locate and extract the central point of each blink event during blink trials.

However, wire movements can also produce peak-like shapes, which poses challenges for this filtering method. While effective in controlled or low-noise environments, the filter is easily triggered by noisy conditions, where artifacts such as wire movements may mimic blink patterns. Figure 1b presents time series segments of both blink and wire movement examples that have been classified as blinks under the current filtering approach, overlaid for comparison. The figure shows that wire movements exhibit greater variability in the regions surrounding the peak, as well as in the overall shape of the peak itself. Current approaches are unable to distinguish between true blinks and wire artifacts, underscoring the limitations of the method in noisier environments.

For each detected peak, baseline values were calculated to provide a reference point for the signal's amplitude. This involved locating the nearest minimum values on either side of the peak by performing binary search with a window size of up to 0.5 seconds in the left and right direction from the peak observed (see algorithm pseudocode in Multimedia Appendix 1). It recursively narrows down the search range to locate a local minimum, while avoiding minor fluctuations.

After establishing the baseline points, we extracted a comprehensive set of amplitude-independent features for each peak. These features include blink duration and various acceleration and velocity metrics, as utilized in prior EOG feature extraction and peak signal analysis studies [32], [33]. A total of 32 peak-related features and label are stored as examples in the dataset, with labels distinguishing natural blinks from noise artifacts.

Figure 1c illustrates examples of EOG signals from two independent singular blink events, with distinct sections of the peak highlighted for clarity. The purple dot at the peak center represents the highest voltage point, detected by the peak detection algorithm. Red dots indicate local maxima in velocity, while blue dots show local acceleration points. Shaded regions in different colors represent key sections of the blink, such as the rising and falling phases, as well as acceleration and deceleration phases. This segmentation captures various aspects of the blink shape, this detailed segmentation provides valuable insights into the blink dynamics, enabling the extraction of relevant blink-related features.

We establish bounds for each feature by discretizing its range into 50 intervals. This discretization splits the feature's values into small, equally spaced segments, enabling a systematic exploration of possible lower and upper bounds that optimize model accuracy. The process begins by identifying the minimum and maximum values of each feature. The range between these values is divided by the bin count (50), yielding an incremental "step size," or delta value, for testing. This delta value determines how much the threshold will shift at each iteration when exploring the bounds. To identify the best lower bound, the algorithm starts from the minimum value and iteratively adds the delta value (e.g., 0.2) to the threshold, testing each increment by culling data points below it and evaluating the model's accuracy with the adjusted dataset. The lower bound with the highest accuracy is selected as the optimal starting point for that feature.

The search then proceeds to find an optimal upper bound, beginning with the maximum value and reducing it by increments of the delta value until reaching the previously identified lower bound. This decremental approach ensures the upper bound remains above the lower bound. Each new threshold is applied to the dataset, and the accuracy is recorded. The upper bound yielding the best accuracy becomes the final threshold for that feature.

The individually optimized lower and upper bounds for each feature are compiled into a list, representing the complete culling thresholds that maximize model performance across the dataset. By discretizing each feature's range into 50 intervals, the *individual search* method ensures a thorough yet efficient exploration of potential thresholds.

## Emotion, EOG, and EDA Monitoring in Cold Pressor Conditions (EMOCOLD) Data Collection

The data collection process employed wearable sensors to record EDA and EOG signals from participants during controlled stress trials. EOG recording used the same setup as the BLINKEO data collection. Electrodes were positioned above and below one eye to detect vertical eye movements by capturing corneo-retinal potential shifts. EDA signals were recorded using a GSR (Galvanic Skin Response) sensor with MCP606 (Microchip Technology) operational amplifiers, operating at an excitation voltage of 0.5V to measure skin conductance. Electrodes were placed on the forehead, chosen for its sensitivity to stress-induced sweat gland activity. The recorded signals were digitized and processed in real-time using an ESP32-S3 WROOM-1 (Espressif Systems) microcontroller, which managed data acquisition, signal processing, and wireless transmission.

Sixteen participants (N=16) between ages 26-31 took part in the study, and demographic information, including race and gender, was collected and is summarized in Table 2a-b. Data was taken from each subject only once. Each trial lasted about 10-15 minutes and was divided into three phases: baseline, CPT (Cold Pressor Test), and recovery. The length of the trial and the data used for feature analysis is as detailed in Table 2c-d.

EOG signals were recorded using a three-electrode configuration designed to capture vertical eye movements, particularly blink activity. Electrodes were positioned as follows: one above the eye, one below the eye, and a reference electrode in the middle of the forehead. This setup effectively captured vertical eye movement signals, with the reference electrode providing signal stability and reducing noise.

For EDA, a single electrode was placed on the forehead to measure changes in skin conductance associated with sympathetic nervous system activation. The forehead was chosen for its accessibility and stable conductance properties, making it suitable for detecting stress-related physiological changes in skin conductance.

Participants wore the device throughout the Cold Pressor Test (CPT) trials, which were conducted to simulate acute stress events. The trials included both physical and environmental stressors. In the cold-water trials, participants immersed their hand in a circulating water bath set to a constant temperature of 0-6°C. Participants maintained immersion for approximately 5 minutes or until voluntary withdrawal. This provided a controlled means of eliciting stress responses.

The design of these trials facilitated the collection of time-series data, capturing participants' physiological reactions to both physical exertion and environmental stressors, thereby providing a comprehensive view of their autonomic responses under varying stress conditions. Features were extracted from partitions of this sensor data, including statistical measures (mean, variance, and standard deviation), signal entropy, peak detection metrics, and frequency-domain characteristics relevant to stress-induced physiological changes. A graphical depiction of the trial methodology can be found in Figure 2.

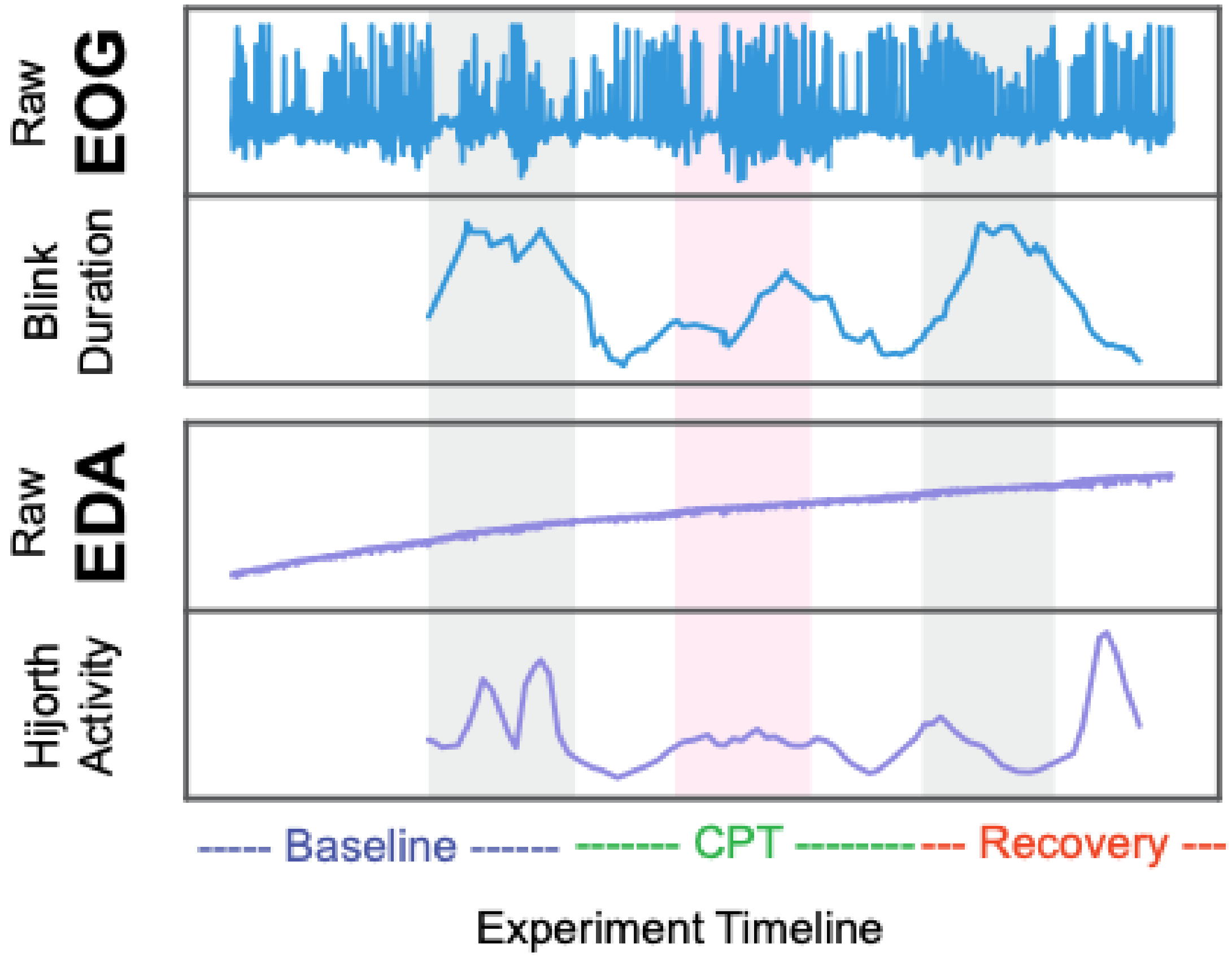

*Figure 2* *This figure presents a visual representation of the experiment timeline and the signals recorded during the experiment, detailing the Baseline, Cold Pressor Test (CPT), and Recovery phases. The raw Electrooculography (EOG) and Electrodermal Activity (EDA) signals across these phases show no immediately clear trend distinguishing the baseline and recovery from the CPT stressor. However, when specific features such as Blink Duration from EOG and Hjorth Activity from EDA are extracted and overlaid, more distinct patterns emerge, and can be used to quantify physiological responses to stress induction and subsequent recovery.*

**Table 2**

| **Characteristic** | **Count** |
|---|---|
| **Assigned Sex** | |
| Male | 11 |
| Female | 5 |
| **Race** | |
| Asian | 11 |
| Hispanic or Latino | 2 |
| White | 1 |
| Middle Eastern or North African | 1 |
| Black or African American | 1 |
| **Total Participants** | 16 |

*Table 2* Characteristics of trials in the EMOCOLD dataset. Demographic details of the study participants, including race and assigned sex.

**Table 3**

| Experiment | Length (Seconds) | | | | |
|---|---|---|---|---|---|
| | Min | 25th Percentile | Median | 75th Percentile | Max |
| **Trial** | | | | | |
| Baseline | 245.6 | 274.0 | 281.7 | 310.0 | 414.8 |
| CPT | 261.9 | 278.4 | 290.4 | 306.4 | 358.0 |
| Recovery | 238.6 | 252.8 | 261.3 | 278.1 | 311.2 |
| **Feature Collection** | | | | | |
| Baseline | 167.5 | 172.1 | 177.0 | 182.3 | 194.0 |
| CPT | 160.6 | 165.0 | 177.2 | 184.1 | 188.2 |
| Recovery | 157.1 | 168.4 | 172.1 | 180.3 | 191.9 |

*Table 3* Summary of trial durations across different experimental phases. Summary of the duration of time EDA and EOG features are collected from, across different experimental phases. For each phase—*Baseline (before hand submersion), Cold Pressor Test (cold water immersion), and Recovery (after hand removal)*—the table lists the minimum, 25th percentile, median, 75th percentile, and maximum duration (in seconds).

At each stage of the experiment—baseline, Cold Pressor Test (CPT), and recovery—participants completed an excerpt of the Positive and Negative Affect Schedule (PANAS) and the State-Trait Anxiety Inventory (STAI-State) to assess their emotional responses. The PANAS measures both positive emotions (e.g., Inspired, Attentive) and negative emotions (e.g., Upset, Nervous) on a 5-point scale, capturing general mood states. The STAI-State survey, consisting of items such as “I feel tense” and “I feel worried”, assesses immediate anxiety levels on a 4-point scale, making it particularly useful for tracking s-anxiety in response to acute stress. The survey recorded at each stage is detailed in the Multimedia Appendix section (Multimedia Appendix 2). Administering these surveys at each stage allowed us to correlate physiological data from EOG and EDA signals with subjective emotional responses, providing a comprehensive view of how participants’ mood and anxiety levels evolved across stress phases.

## Electrooculography (EOG) Signal Segmentation

In analyzing Electrooculography (EOG) signals, we segmented the data to isolate individual blink peaks, which are essential for understanding blink dynamics in response to stress. From these peaks, we extracted 35 features, including blink duration, amplitude, frequency, and various acceleration and velocity metrics. A comprehensive list of these features and their definitions is provided in Multimedia Appendix 4.

### Electrodermal Activity (EDA) Signal Segmentation

The tonic and phasic components of skin conductance reveal different aspects of autonomic arousal, with the tonic level representing a stable baseline and the phasic response capturing transient, stimulus-driven changes. Tonic signals vary significantly across individuals due to factors like skin type and hydration, making them challenging to analyze consistently in relation to specific stress events. Phasic responses, however, reflect rapid fluctuations in skin conductance directly tied to acute stress or anxiety-inducing stimuli, characterized by quick rises and gradual declines.

Phasic signals were divided into rise and fall phases to capture the dynamics of the skin conductance response, which is indicative of sympathetic nervous system activation. Specifically, peaks were detected by identifying rapid increases in skin conductance (rise phases) followed by gradual decreases (fall phases). To preprocess the EDA data and extract the phasic signal, motion artifacts were identified and removed, to make the data suitable for downstream features. A first-order low-pass Butterworth filter was applied to isolate low-frequency components indicative of meaningful physiological signals.

This signal was divided into windows of 1 second in length. Each section was analyzed to determine key features, such as mean value, signal range, and standard deviation. 15 features were extracted from these windows, and the full list of features and their definitions can be found in Multimedia Appendix 3. These features are critical for quantifying the intensity and duration of autonomic arousal events, providing valuable insights into stress response dynamics. The segmentation process allowed for the extraction of detailed temporal characteristics of each skin conductance event, facilitating a comprehensive analysis of physiological arousal under stress.

## Results

### Blink Identification EOG (BLINKEO) Analysis

Building upon the non-intentional blink signal processing outlined by previous research [34] [35], a feature bounding analysis aligned closely with the article's approach of differentiating blink events based on slope and derivative features. By using blink duration alone as a feature, we achieved a classification accuracy of 87.46% and an F1 score of 0.7999 in distinguishing blinks from wire movements (see Multimedia Appendix 5). This suggests that feature extraction can yield strong performance metrics. Even without deep learning techniques, finding the right markers of blink peaks can reach the same efficacy of the article's outlined slope-based signal differentiation.

In our approach, we systematically evaluate all possible combinations of five selected features to optimize classification performance for distinguishing blink events from wire movements. For each feature combination, we apply a breadth-first search (BFS) traversal to explore and fine-tune the upper and lower bounds of each feature, seeking the configuration that maximizes classification accuracy.

The BFS traversal begins with initializing the bounds for each feature to cover its entire observed range, ensuring that no data points are culled at the outset. Each feature range is discretized into 15 bins, allowing for incremental adjustments to the bounds with a step size (delta) calculated as the range divided by the number of bins. These initial bounds are stored as a “node” in the BFS queue, representing a unique culling configuration.

During each iteration of the BFS traversal, we dequeue a culling configuration and calculate its classification accuracy and F1 score using a performance function. If the configuration achieves a higher accuracy than previously recorded, it becomes the current optimal configuration. The BFS traversal then generates neighboring configurations by slightly tightening the bounds for each feature—either increasing the lower bound or decreasing the upper bound by the computed delta. Each of these neighboring configurations, if unvisited, is added to the queue for further exploration.

This BFS traversal continues until all relevant bound configurations for the current feature combination are evaluated. The outcome is an empirically derived set of feature bounds that maximizes classification performance for each combination of features. By applying this process across all combinations of the selected five features, we ensure a comprehensive search of the parameter space, yielding an optimal culling pipeline tailored for precise blink detection. This method demonstrates the robustness of combining BFS with multi-feature analysis to achieve a high-performing, data-driven classification model.

In our approach, we select combinations of five high-quality features and use a breadth-first search (BFS) traversal to optimize their combined bounds for maximal classification performance. For each combination, BFS systematically explores adjustments to the upper and lower bounds of each feature, identifying the optimal configuration that yields the highest accuracy and F1 score.

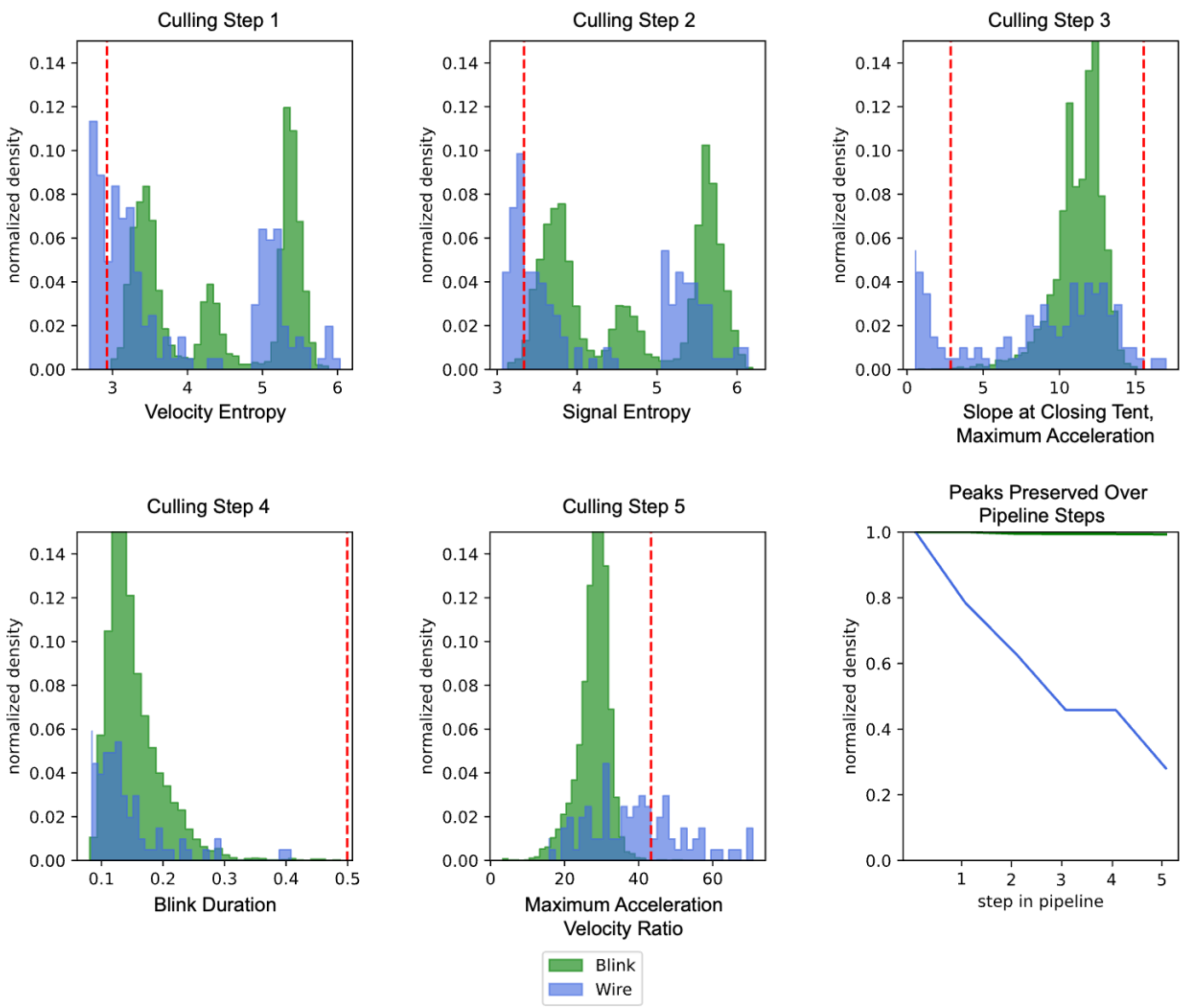

*Figure 3* Optimal Culling Steps for Differentiating Blink Events from Wire Movement Artifacts in EOG Data. This figure presents the sequential culling steps optimized to achieve the highest accuracy and F1 score in distinguishing blink events (green) from wire artifacts (blue) in EOG data. Each subplot demonstrates a unique culling step, applying specific feature thresholds to progressively refine the data. The final subplot, "Peaks Preserved Over Culling Pipeline," illustrates the proportion of retained peaks at each stage for both blink and wire signals, showcasing the efficacy of each step in isolating genuine blink events.

The optimal feature combination achieved an accuracy of 98.17% and an F1 score of 0.8734, utilizing five key features that capture distinctive characteristics of blink dynamics. These features include: *Velocity* Entropy, the entropy of the first derivative of the signal, which measures the variability and complexity of the blink motion; *Signal* Entropy, the entropy of the signal itself, providing a broader assessment of the overall blink pattern; *Slope at Closing Tent, Maximum Acceleration*, the maximum acceleration during the closing phase of a blink, which isolates the rapid deceleration typical of blink closure; *Blink Duration*, representing the total time span of the blink event; and *Maximum Acceleration Velocity Ratio*, the ratio between the maximum acceleration and maximum velocity during the closing phase, which captures the relationship between these peak dynamics, indicative of voluntary eye closure. The results of each feature bounding step, against the BLINKEO labelled examples, is shown in Figure 3.

These features together form a comprehensive representation of blink characteristics, enabling differentiation of blinks from other signal types in the culling pipeline. This highlights how strategically selected bounds on multiple features, when combined, result in high classification performance without relying on complex algorithms.

## Emotion, EOG, and EDA Monitoring in Cold Pressor Conditions (EMOCOLD) Analysis

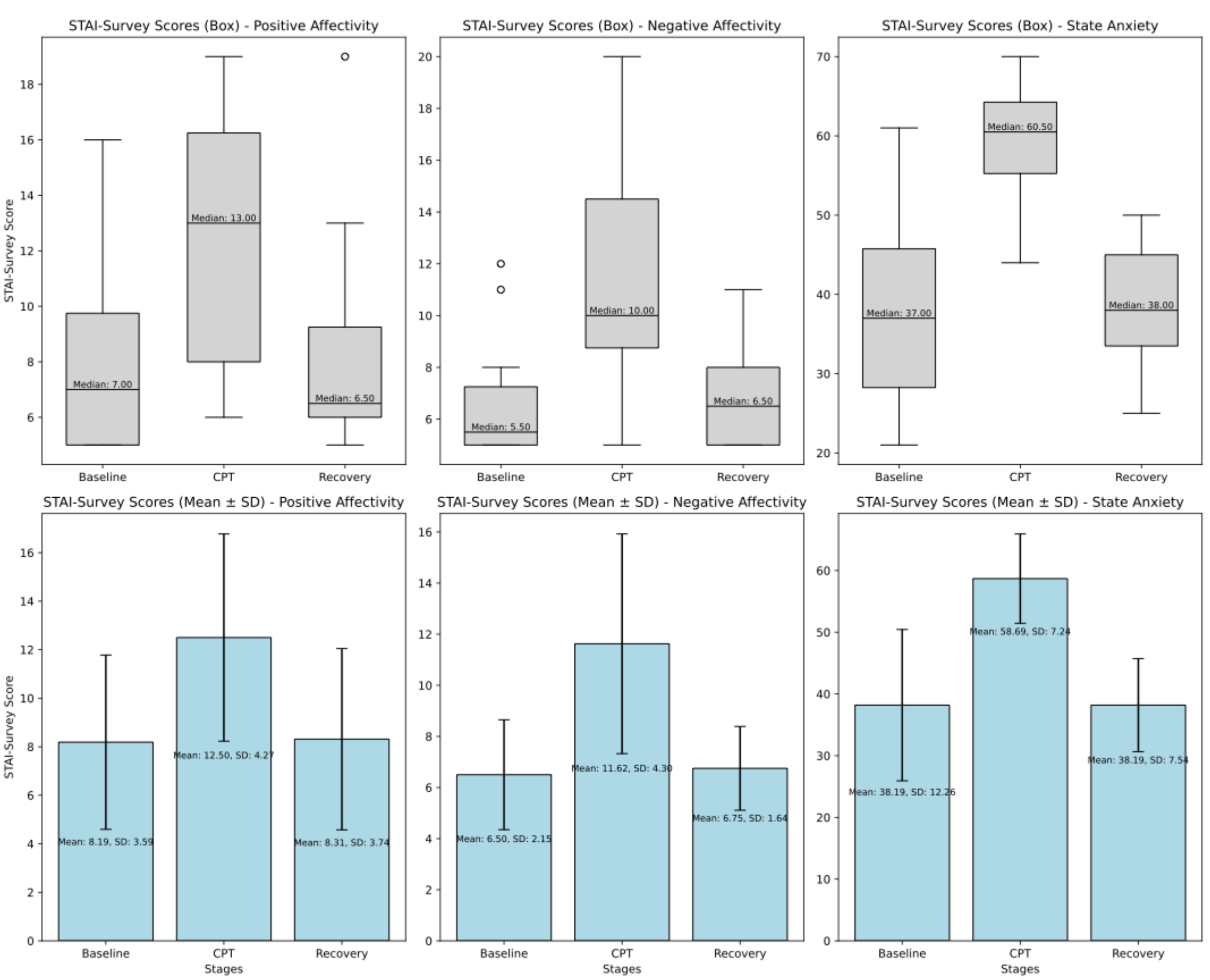

*Figure 4 User-reported survey responses during each stage of the trial, displaying both box-and-whisker plots and column graphs for Positive Affectivity, Negative Affectivity, and State Anxiety (S-Anxiety) across the Baseline, CPT, and Recovery stages. During CPT, participants showed higher levels of positive affectivity, negative affectivity, and stage anxiety. Elevated levels recovered to baseline responses when participants took their hand out of the cold-water bath during the recovery phase.*

The EMOCOLD dataset analysis highlights significant physiological and emotional responses to acute stress induced by the Cold Pressor Test (CPT). Figure 4 shows participants' aggregated self-reported survey scores for positive affectivity, negative affectivity, and s-anxiety across the three trial stages: Baseline, CPT (Cold Pressor Test), and Recovery. In figure 4, for each stage, survey responses were summarized and visualized using box plots, which display the distribution of scores. Positive Affectivity and Negative Affectivity are scored on a scale of 5-25, and State Anxiety is scored on a scale of 20-80.

Participants reported increased positive and negative affectivity, as well as elevated s-anxiety during CPT, which returned to baseline during recovery. This dual affective response suggests heightened arousal may include both alertness and discomfort. The recovery phase indicates effective autonomic regulation, as emotional states normalized once the stressor was removed. These findings validate CPT as a method for inducing short-term anxiety.

## SHAP (SHapley Additive exPlanations) Analysis

SHAP (SHapley Additive exPlanations) analysis is a method used to explain the output of machine learning models by breaking down the prediction into contributions from each feature. SHAP values are based on Shapley values from cooperative game theory, which attribute the impact of each feature on the model's output by treating each feature as a "player" in a game and calculating its contribution to the final prediction.

In this study, SHAP analysis was performed on combinations of five features, selected from the total feature set of 15 EDG and 35 EOG features, highlighting the significance of how certain biomarkers, used together, reveal more prominent interactions and effects on model predictions. This approach underscores that certain biomarkers, while potentially less impactful individually, can demonstrate substantial importance when analyzed as part of a group. By evaluating these interactions, we understand how combinations of features can provide insights into the model's behavior that single-feature analyses might overlook.

The quality of a set of features is determined by considering their collective contribution to the model's predictions, measured through the mean absolute SHAP values across the dataset. A high-quality set of features is one where the combination of features demonstrates substantial importance, as indicated by a higher mean absolute SHAP values. This benchmark reflects not only the magnitude of individual contributions but also the degree to which the features, as a group, interact to enhance the predictive power of the model.

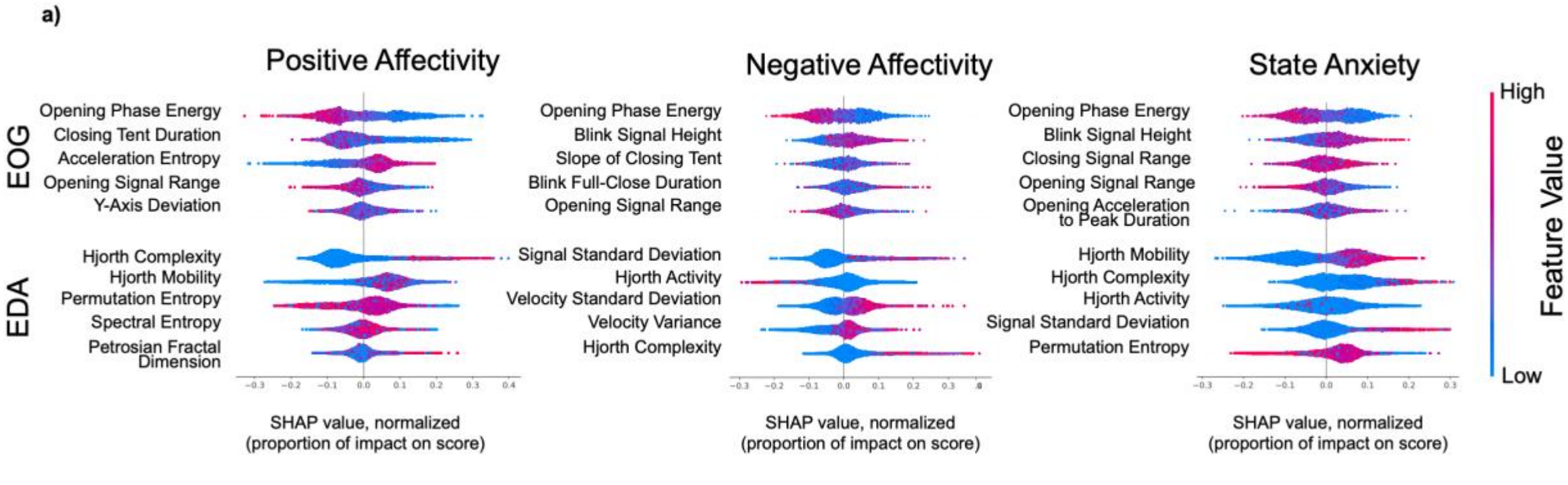

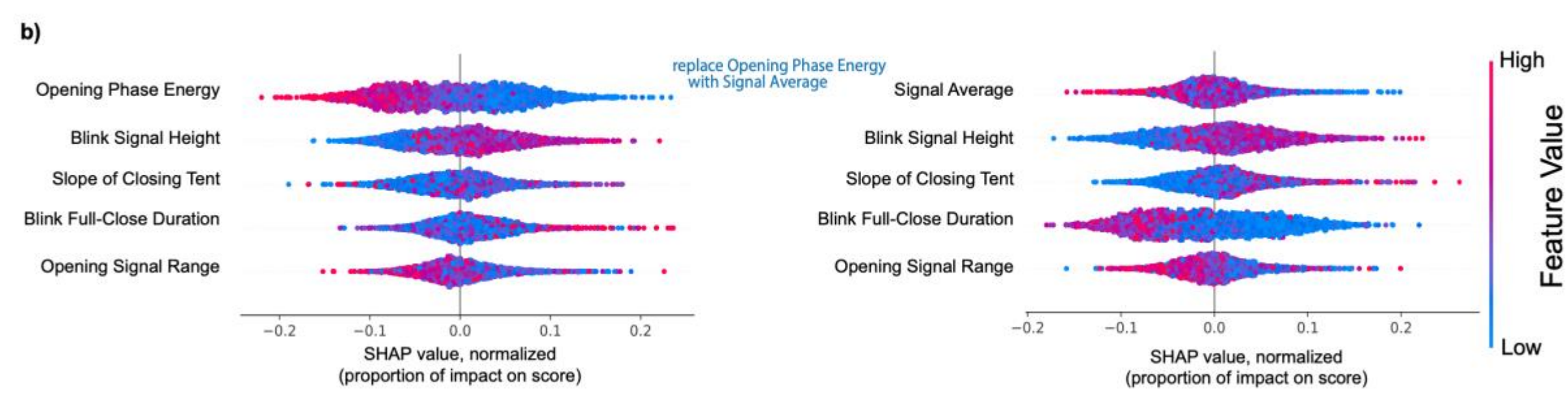

*Figure 5 5a*. SHAP Analyses for optimal combinations of five EOG Features (top row) and five EDA Features (bottom row), for Positive Affectivity (left column), Negative Affectivity (middle column), and State Anxiety (right column). *5b*. SHAP Analysis of Feature Combinations. This analysis explores the quality of distinguishing different affectivity levels using different sets of features. This is an example of five EOG features and their impact on the Negative Affectivity score. Substituting one key feature with another can reveal new interdependencies among remaining features, thereby enhancing the model's interpretability.

The SHAP value maps provide insights into how various EOG features used in combination, and EDA features used in combination, contribute to predictions for positive affectivity negative affectivity, and s-anxiety. Each SHAP sub-plot illustrates the impact of individual features on model outputs, with higher SHAP values (toward the right) signifying a positive contribution to the prediction, and lower SHAP values (toward the left) indicating a negative contribution. Figure 5a highlights the SHAP analysis identifying the combination of features that best polarize model predictions across the affective states.

### *Electrooculography (EOG) Feature Analysis*

Among the EOG features analyzed, the *Opening Phase Energy*, the integral of the opening phase of the peak signal, and *Opening Signal Range*, the amplitude of the opening phase of the peak signal, consistently appeared in optimal feature combinations across all three outputs, suggesting their robustness as predictors. Additionally, the *Signal Height* feature exhibited a particularly strong influence on predictions for negative affectivity and s-anxiety, underscoring its significance in these contexts.

### *Electrodermal Activity (EDA) Feature Analysis*

Among the EDA features analyzed, Hjorth parameters and the *Signal Standard Deviation* emerged as important predictors across the different affective states. These findings highlight the importance of analyzing feature interactions to reveal critical combinations that drive model performance, offering deeper insights into the physiological signals underpinning emotional and stress-related states.

The SHAP analyses in Figure 5b illustrate the importance of considering features in combination when identifying the most relevant biomarkers. By selecting sets of five features, we aim to identify a group of biomarkers that not only are individually relevant but also work effectively together. In Figure 5b, the inclusion of the feature *Opening Phase Energy* contributes significantly to the model's performance, yielding a well-defined distinction in SHAP values. When *Opening Phase Energy* is removed from the features considered, model performance decreases, and features such as *Blink Full-Close Duration* appear to show more distinction.

## Discussion

The main findings of this study show the potential of electrooculography (EOG) and electrodermal activity (EDA) as powerful tools for identifying nuanced physiological biomarkers associated with state anxiety. Through the development and analysis of the BLINKEO and EMOCOLD datasets, we have introduced novel datasets and used advanced feature extraction techniques with interpretability methods such as SHAP analysis to uncover anxiety-specific markers. Our results emphasize the importance of understanding biomarkers in their context-dependent interactions and collective contributions to predictive models.

By systematically evaluating combinations of features, we mitigated challenges often faced in the literature, where biomarkers show inconsistent or non-significant correlations with anxiety due to situational variability. For instance, while blink rate and skin conductance metrics have been previously explored, our analysis reveals that their predictive utility depends heavily on contextual factors, such as the type and intensity of the stressor. For example, biomarkers like blink duration and skin conductance peaks performed well under controlled Cold Pressor Test (CPT) conditions but may not generalize to other stress-inducing scenarios like public speaking. This underscores the need for adaptive, context-sensitive models that account for the situational variability of physiological responses.

A key contribution of this work is the identification of feature combinations that consistently provide reliable predictions. For EOG data, features like blink duration, peak height, and the opening integral were shown to be robust predictors across various emotional states. Similarly, for EDA data, features such as the mean signal, permutation entropy, and Hjorth activity emerged as significant contributors. By leveraging SHAP analysis, we identified not only which features are most relevant but also when and how they interact to enhance model performance. This approach offers a more comprehensive understanding of physiological responses compared to studies focusing solely on single-feature analyses.

Our findings bridge a critical gap in the literature by offering a systematic approach to addressing the variability and context-dependence of physiological biomarkers. This research advances the field by providing a framework for building more robust, interpretable, and context-sensitive models for anxiety assessment. The ability to dynamically adapt to different stress scenarios makes these biomarkers more applicable to real-world settings, paving the way for more personalized and effective mental health interventions.

## Limitations

This study advances state anxiety biomarker detection using Electrooculography (EOG) and Electrodermal Activity (EDA), but several limitations should be noted. The participant pool (N=16) was demographically skewed, with a predominance of male and Asian participants, limiting generalizability. Data was collected only once per subject, preventing analysis of intra-individual variability over time. Future studies should incorporate larger and more diverse populations with longitudinal data.

The Cold Pressor Test (CPT) was conducted in a controlled lab environment, which may not fully reflect real-world anxiety triggers. Additionally, motion artifacts in EOG recordings, despite filtering efforts, could impact signal clarity. EDA signals were recorded using a single forehead electrode, though different placements (e.g., fingertips) may improve accuracy. Improved artifact detection and additional motion-tracking sensors could enhance data quality.

Feature selection for SHAP analysis focused on optimizing interpretability, but alternative selections may yield different insights. Models and analyses constructed using this dataset may not generalize well to other stress-inducing scenarios. External validation using independent datasets is necessary to confirm these findings.

## Future Work

Future work should focus on validating these findings across diverse populations and stress-inducing contexts to further enhance the generalizability of these biomarkers. An important next step is to investigate potential gender-based and race-based differences in physiological responses to acute stress and our current methods of inducing stress, as our current study was not explicitly designed for such analysis but acknowledges its relevance. Additionally, integrating these models into wearable technology has the potential to revolutionize mental health monitoring, providing real-time, personalized insights that could transform how we understand and manage anxiety. By addressing the challenges of situational variability and leveraging the strengths of combined biomarker analyses, this study contributes significantly to the growing field of wearable health technology and its applications in mental health.

# Acknowledgements

## Funding Statement

This work was funded by the Translational Research Institute for Space Health through NASA NNX16AO69A, Office of Naval Research grant nos. N00014-21-1-2483 and N00014-21-1-2845, Army Research Office grant no. W911NF-23-1-0041, National Institutes of Health grant nos. R01HL155815 and R21DK13266, National Science Foundation grant no. 2145802, National Academy of Medicine Catalyst Award and High Impact Pilot Research Award no. T31IP1666 from the Tobacco-Related Disease Research Program and Heritage Medical Research Institute (all to W.G.). T.K.H. acknowledges the support from National Institutes of Health grant nos. T32HL144449 and T32EB027629.

## Human and Animal Research Statements

This study was conducted in accordance with ethical guidelines for research involving human participants. 16 participants were recruited following established ethical guidelines as delineated in protocols approved by the institutional review board at the California Institute of Technology (Caltech) (protocol numbers IR22-1280 and IR21-1102). Participants were not compensated. Participants were screened based on specific exclusion criteria, including non-English speakers unable to understand survey requirements, inability to provide informed consent, medication use affecting psychiatric states, pregnancy, irregular eye conditions (e.g., ocular dysmetria), and pre-existing psychiatric or physical illnesses (e.g., depression, anxiety, hypertension, hyperlipidemia, or chronic cardiovascular disease). All participants' data was fully anonymized, with identifying information removed and data transmission secured using byte-splicing encryption methods. The study adhered to data privacy and security protocols to ensure the confidentiality and protection of participants.

# Data Availability

All meta-datasets are publicly available and can be requested with the proper EULA form from the proprietor, as noted within each respective published works.

The BLINKEO and EMOCOLD datasets can be accessed at:
BLINKEO: https://github.com/jadee-dao/stress-biomarkers-public-dataset/tree/main/blinkeo
EMOCOLD: https://github.com/jadee-dao/stress-biomarkers-public-dataset/tree/main/emocold

## Multimedia Appendix Captions

Multimedia Appendix 1: The bounds of each peak were determined by performing two binary searches within the position domain of the signal—one to the left of the peak and one to the right.

Multimedia Appendix 2: The survey items from the Positive and Negative Affect Schedule (PANAS) and the State-Trait Anxiety Inventory (STAI-State) were used to assess participants' emotional and anxiety responses during the experiment. The PANAS scale consists of 10 items measuring Positive Affectivity and Negative Affectivity, each rated on a 1-5 Likert scale, where higher scores indicate stronger affective states. The STAI-State consists of 20 items assessing state anxiety, measured on a 1-4 Likert scale, where responses indicate varying degrees of agreement with statements reflecting anxiety levels. Higher scores in negative affectivity and anxiety-related items indicate greater distress, while higher scores in positive affectivity items indicate greater emotional well-being. The table below details each item, its corresponding scale, and the affectivity or anxiety dimension it evaluates.

Multimedia Appendix 3: Table of feature names and definitions extracted from windowed segments of EDA signals.

Multimedia Appendix 4: Table of feature names and definitions extracted from windowed segments of EOG signals.

Multimedia Appendix 5: A breadth-first search was performed to find the optimal range for distinguishing blinks from noise artifacts using the Blink Duration feature, which was extracted from EOG signal peaks using our method. The identified bounds of 0.1227–0.3990 seconds align with values reported in the literature.

## Conflict of Interest

Authors declare no competing interests.

## Author Contributions

S.A.S. and J.D. conceived the project. J.D. and S.A.S. led the main study, collected the overall data, and contributed to the data analysis. R.L. and S.L.S. contributed to the platform development, characterization, human studies, and data processing. J.D. and S.A.S. co-wrote the paper. All authors provided feedback on the manuscript.

## Abbreviations

BFS – breadth-first search
BLINKEO – Blink Identification EOG Dataset
CPT – cold pressor test
EDA – electrodermal activity
EMOCOLD – Emotion, Electrooculography, and Electrodermal Activity Monitoring in Cold Pressor Conditions Dataset
EOG – electrooculography
SHAP – SHapley Additive exPlanations

## Generative AI

Generative AI [ChatGPT, version GPT-4o, OpenAI, 2025] was only used to assist with grammar correction and formatting of the author's original text for this manuscript. These tools were employed to improve clarity while preserving the author's original ideas and content. All AI-assisted outputs were thoroughly reviewed and edited by the author for accuracy and consistency prior to submission.